\documentclass[prl,aps,showpacs,twocolumn,amsmath,amssymb]{revtex4-1}
\usepackage{graphicx}
\usepackage{color}

\begin{document}


\title{Realization of a Resonant Fermi Gas with a Large Effective Range}


\author{E. L. Hazlett, Y. Zhang, R. W. Stites, and K. M. O'Hara}
\affiliation{Department of Physics, Pennsylvania State University,
University Park,\nolinebreak \,Pennsylvania 16802-6300, USA}


\date{\today}

\begin{abstract}
We have measured the interaction energy and three-body recombination rate for a two-component Fermi gas near a narrow Feshbach resonance and found both to be strongly energy dependent.  Even for deBroglie wavelengths greatly exceeding the van der Waals length scale, the behavior of the interaction energy as a function of temperature cannot be described by atoms interacting via a contact potential.  Rather, energy-dependent corrections beyond the scattering length approximation are required, indicating a resonance with an anomalously large effective range.  For fields where the molecular state is above threshold, the rate of three-body recombination is enhanced by a sharp, two-body resonance arising from the closed-channel molecular state which can be magnetically tuned through the continuum.  This narrow resonance can be used to study strongly correlated Fermi gases that simultaneously have a sizeable effective range and a large scattering length.
\end{abstract}



\maketitle

Experimental studies of unitary Fermi gases have exclusively used energetically broad Feshbach resonances to realize strong interactions~\cite{Thomas02B,Salomon03,Jin04,Ketterle04,Grimm04,Ketterle05,Thomas05A,Thomas05B,Hulet06,Jin06,Thomas07,Grimm08B,Thomas09,Salomon10,Mukaiyama10,Grimm11B}.  For broad resonances, which exhibit negligible energy dependence over the scale of the Fermi energy $E_F$, pairwise interactions in a two-component Fermi gas can be completely parameterized by the scattering length, $a$~\cite{Gurarie07,Zwierlein08,Tiesinga10}.  Thus, the equation of state (EOS) of the Fermi gas in the unitary limit ($\left|a \right| \rightarrow \infty$) must be a universal function (dependent only on $E_F$ for each state and the reduced temperature) that describes any dilute Fermi gas with resonant, zero-range (i.e. energy-independent) interactions~\cite{Ho04,Mueller04,Salomon10,Mukaiyama10}.

More generally, however, strongly-interacting Fermi systems can also result from resonant, short-range interactions that are energy dependent and this energy dependence can significantly alter the collective behavior~\cite{Schmidt03,Kokkelmans04,Pethick04}.  At minimum, the EOS in the limit  $\left| a \right| \rightarrow \infty$ now depends on an additional parameter, the effective range $r_{\mathrm{eff}}$, which characterizes the first-order term in the low-energy expansion of the $s$-wave phase shift $\delta$ (i.e. $k\, \cot \delta = -1/a + \frac{1}{2} r_{\mathrm{eff}} \, k^2 + ...$).  This correction is pertinent to neutron matter at densities in neutron stars for which $r_{\mathrm{eff}} \, k_F \sim 1$ where $k_F$ is the Fermi wavenumber~\cite{Pethick05}.
In other cases where $r_{\mathrm{eff}} < 0$, theoretical estimates suggest that the unitary Fermi gas becomes even more strongly interacting as $\left| r_{\mathrm{eff}} \right|$ is increased from zero~\cite{Kokkelmans04,Pethick05,Ho11}, potentially yielding superfluids with the highest critical temperature ever achieved.  Most dramatically, the very existence of certain novel phases of matter (e.g. the breached-pair superfluid phase~\cite{Wilczek05}) depend on interactions having a particular momentum dependence.

Narrow Feshbach resonances (FR) should allow for the study of such behavior in an atomic Fermi gas as they have been predicted to exhibit a strong energy dependence over the scale of $E_F$~\cite{Pethick04,Kokkelmans04,Kokkelmans06,Gurarie07,Kokkelmans08,Zwierlein08,Tiesinga10}. The effective range $r_{\mathrm{eff}}$ in the vicinity of a FR is inversely proportional to its width and, for narrow resonances, can be large in comparison to $1/k_F$ or even the van der Waals length $\ell_{\mathrm{vdW}}$~\cite{Tiesinga10}, potentially allowing for the study of dilute Fermi gases with large $a$ and large $r_{\mathrm{eff}}$.

In this Letter, we experimentally demonstrate that interactions in a two-component Fermi gas near a narrow FR cannot simply be parameterized by a field-dependent scattering length but require an energy-dependent scattering length or, near the resonance, an anomalously large effective range.  This is done by comparing the measured interaction energy for a two-component Fermi gas at different temperatures with predictions of mean-field theory.  The interaction energy is measured by radio-frequency spectroscopy using a third state~\cite{Ketterle03A,*Jin03B}.  Measurements with cold gases accurately determine the location and width of the resonance.  For hotter samples, the field-dependence of the interaction energy exhibits an asymmetric lineshape which we interpret to be a signature of energy dependent interactions.  We further show that inelastic loss from the gas by three-body recombination is enhanced by a sharp resonance in the two-body continuum as the molecular state is tuned above threshold.  From the inelastic loss data we can extract the vibrational relaxation rate constant due to atom-dimer collisions for the molecular state associated with the resonance.

\begin{figure}
\begin{center}
\includegraphics[width=3.35in]{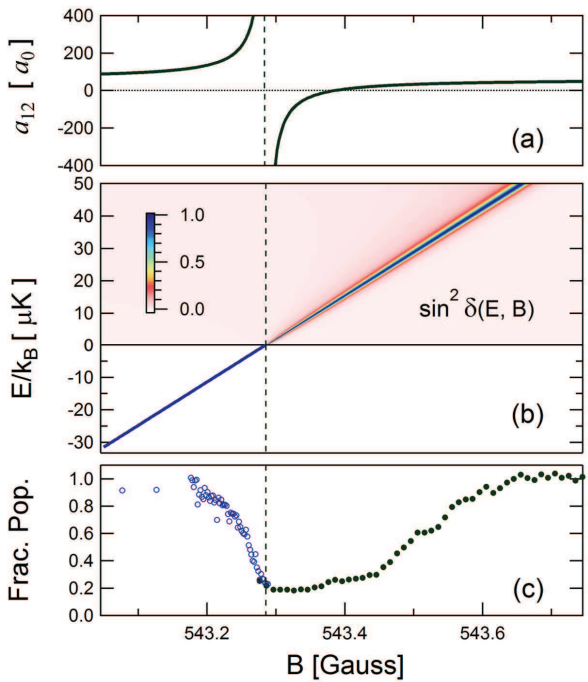}
\end{center}
\caption{\label{fig:NarrowRes} Narrow Feshbach resonance in $^6$Li.  (a) The scattering length as a function of magnetic field.  (b)  For $E < 0$, the binding energy of the molecular state relative to threshold is shown.  The molecular state crosses threshold at $B_\infty = 543.286(3)\,~\mathrm{G}$.  For $E > 0$, a contour plot of $\sin^2 \delta(E,B)$ demonstrates the sharp resonance that the molecular state creates as it is tuned through the continuum. (c) Fraction of atoms remaining after a 100~ms hold time at various $B$-fields when the resonance is approached from below ($\circ$) or above ($\bullet$) showing an asymmetric loss resonance for a cloud at $T = 7\,\mu{\mathrm{K}}$.}
\end{figure}

In our experiments, we investigate the narrow resonance which occurs in $s$-wave collisions between the two lowest-energy hyperfine states in $^6$Li (states $\left| 1 \right\rangle$ and $\left| 2 \right\rangle$) near a field of $B_{\infty} = 543.3\,~\mathrm{G}$~\cite{Thomas02A}.  Long-lived $^6$Li$_2$ molecules had previously been created using this resonance~\cite{Hulet03}.  In the vicinity of the resonance, the scattering length varies as $a(B) = a_{\mathrm{bg}} \left(1 - \Delta/(B - B_{\infty}) \right)$ (see Fig.~\ref{fig:NarrowRes}(a)).  Here, $\Delta = 0.1\,~\mathrm{G}$ is the resonance width and $a_{\mathrm{bg}} \simeq 62\,a_0$ is the background scattering length which varies relatively slowly with field.  This narrow FR occurs when the least bound vibrational level of the $X\,^1\Sigma_g^+$ singlet potential crosses threshold.  The energy of the molecular state relative to threshold varies as $E_{c} = \mu_r \,(B - B_{\infty})$ where $\mu_r$ is the difference between the magnetic moment of the colliding atoms and the bare molecular state given in this case by $\mu_r = 2 \, \mu_B$ where $\mu_B$ is the Bohr magneton.

Near threshold, the $s$-wave scattering phase shift $\delta$ is
\begin{eqnarray}
\delta = \delta_{\mathrm{bg}} - \tan^{-1}\left(\frac{\Gamma/2}{E - E_{c}} \right),
\label{eqn:DeltaRes}
\end{eqnarray}
the sum of a background phase shift $\delta_{\mathrm{bg}} = - k\,a_{\mathrm{bg}}$ and a resonant phase shift with a Breit-Wigner form~\cite{Tiesinga10}.
Resonance occurs when the relative kinetic energy $E = \hbar^2 k^2/m$ is equal to the energy of the molecular state relative to threshold $E_{c}$.  Here, $\hbar\,k$ is the relative momentum and $m$ is the mass of a $^6$Li atom.  Near threshold, the energy dependent resonance width $\Gamma(E) = 2\, k \, a_{\mathrm{bg}} \, \mu_r \, \Delta$ describes the finite lifetime of the bound state due to coupling to the continuum.  A contour plot of $\sin^2 \delta$ in the upper half of Fig.~\ref{fig:NarrowRes}(b) illustrates the sharp resonance that tunes through the continuum when the molecular state is above threshold.

The phase shift in Eq.~\ref{eqn:DeltaRes} determines the scattering amplitude $f = 1/(k\,\cot\delta - i k)$ where, as in Ref.~\cite{Ho11}, we find:
\begin{eqnarray}
k\,\cot\delta & = &   - \frac{1}{a_{\mathrm{bg}}} \, \frac{\frac{\hbar^2 k^2}{m} - \mu_r \, (B - B_\infty)}{\frac{\hbar^2 k^2}{m} - \mu_r \, (B - B_\infty) + \mu_r \, \Delta},
\label{eqn:KCotDelta}
\end{eqnarray}
assuming that $k\,a_{\mathrm{bg}} \ll 1$.  Note that near the resonance (i.e. for $(B - B_\infty) \ll \Delta$) and near threshold, the scattering phase shift is given by the effective range expansion, $k \cot\delta = -\frac{1}{a_{\mathrm{res}}} + \frac{1}{2}\,r_{\mathrm{eff}}\,k^2$ where $a_{\mathrm{res}} = a_{\mathrm{bg}} \left(\frac{\Delta}{B_\infty - B} \right)$ and $r_{\mathrm{eff}} = - \frac{2\,\hbar^2}{m} \frac{1}{a_{\mathrm{bg}} \mu_r \Delta}$.  For the narrow FR studied here, the effective range, being inversely proportional to $(a_{\mathrm{bg}}\,\Delta)$, has an anomalously large value $r_{\mathrm{eff}} = - 7 \times 10^4\,a_0$ with a magnitude much larger than the average interparticle spacing in the dilute Fermi gases studied here.  Narrow FRs in $^6$Li-$^{40}$K mixtures should also exhibit $\left| r_{\mathrm{eff}} \right| \gg \ell_{\mathrm{vdW}}$, though the associated energy dependence has not been verified~\cite{Grimm08A,Dieckmann10}.

To study the narrow FR in $^6$Li we have (1) monitored the loss of atoms as a function of magnetic field, and (2) performed radio-frequency (rf) spectroscopy to measure interaction energy shifts using the transition between state $\left| 2\right\rangle$ and the third lowest energy hyperfine state in $^6$Li (state $\left| 3  \right\rangle$).  We can prepare degenerate or thermal mixtures of $^6$Li atoms with equal populations in states $\left|1 \right\rangle$ and $\left| 2 \right\rangle$ by evaporative cooling in a single or crossed-dipole trap~\cite{OHara09}.  To control the density and temperature of atoms in our experiments, we adjust the degree of evaporative cooling and the final trap depth following evaporation.

Our initial observation of the narrow FR is shown in Fig.~\ref{fig:NarrowRes}(c) which plots the fraction of atoms remaining in the optical dipole trap after a 100~ms hold time at each field.  For this data, the average density per spin state $n \sim 10^{13}\,{\mathrm{cm}}^{-3}$, the temperature $T \simeq 7\,\mu{\mathrm{K}}$, and the $\circ$'s ($\bullet$'s) show the fraction remaining when the FR is approached from a field well-below (well-above) resonance at a ramp-rate $\simeq 50\,{\mathrm{mG}}/{\mathrm{ms}}$.   Atom loss from this system must be due to three-body recombination since two-body inelastic processes are frozen out at this temperature for a mixture of $^6$Li atoms in states $\left| 1 \right\rangle$ and $\left| 2 \right\rangle$.  This feature is asymmetric with loss preferentially occurring on the high-field (atomic) side, in contrast to the behavior observed for Fermi gases near broad resonances~\cite{Ketterle02}.  We show that the asymmetry results from the fact that the molecular state enhances three-body recombination as it tunes through the continuum.  Higher-temperature gases can have collisions sufficiently energetic to satisfy the resonance condition $E = \mu_r \, (B - B_\infty)$ even for magnetic fields far-detuned above resonance (i.e. for $k_B\,T \sim \mu_r\,(B - B_\infty) \gg \mu_r\,\Delta$).  Thus, the width of the loss feature on the high-field side of resonance grows as the temperature is increased.

\begin{figure}
\begin{center}
\includegraphics[width=3.25in]{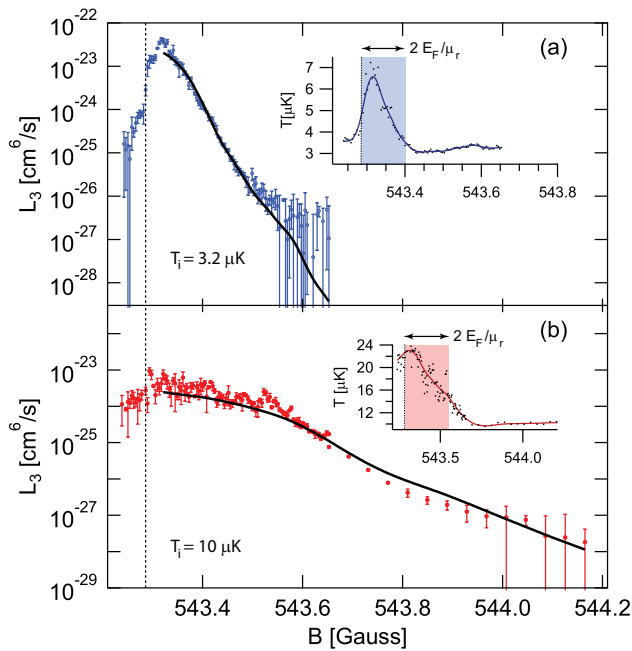}
\end{center}
\caption{\label{fig:InelasticLoss} Recombination loss rates for gases initially prepared at (a) $T_i = 3.2\,\mu{\mathrm{K}}$ and (b) $T_i = 10\,\mu{\mathrm{K}}$.  The solid lines are single-parameter fits to a model (Eq.~\ref{eqn:L3}) which assumes that the molecular state resonantly enhances loss as it is tuned through the continuum.}
\end{figure}

To investigate the dependence of loss on temperature and field near this FR, we monitor the number of atoms as a function of time at fixed fields with samples initially prepared above resonance at one of two different temperatures.  The density decays according to $\dot{n} = - L_3 \, n^3$ due to three-body recombination where $L_3$ is the three-body recombination loss rate constant.  For initial $T_i = 3.2\,\mu{\mathrm{K}}$ ($T_i = 10\,\mu{\mathrm{K}}$), we have extracted $L_3$ for different $B$-fields as shown in Fig.~\ref{fig:InelasticLoss}(a) (Fig.~\ref{fig:InelasticLoss} (b)).  In both cases, the sample is initially at $T \simeq 0.5 \, T_F$, allowing us to analyze this weakly degenerate system as a thermal gas~\cite{OHara09}.  As the $B$-field approaches resonance, the temperature of the cloud increases as shown in the Fig.~\ref{fig:InelasticLoss} insets.  A smoothing spline is fit to the temperature data (solid lines in Fig.~\ref{fig:InelasticLoss} insets) and used to compute $n$ and extract $L_3$~\footnote{The geometric mean of the trap frequencies for the $3.2\,\mu{\mathrm{K}}$ ($10\,\mu{\mathrm{K}}$) data is $\omega = 2\,\pi \times 1.3(1)\,{\mathrm{kHz}}$ ($\omega = 2\,\pi \times 3.6(4)\,{\mathrm{kHz}}$) and there are initially $N \simeq 3(1)\times 10^5$ ($N = 1.7(5) \times 10^5$) atoms in each spin state.}.

To understand the behavior of $L_3$ above resonance, we consider the fact that two atoms with relative kinetic energy $E \simeq E_c$ are resonantly coupled to the vibrationally excited molecular state responsible for the FR which can decay in a subsequent atom-dimer collision.  A narrow FR which couples two colliding atoms to a resonant state which has a finite lifetime ($\hbar/\Gamma_0$) due to inelastic decay, induces loss with a two-body loss rate coefficient~\cite{Clark09}
\begin{eqnarray}
L_2 = \frac{h}{m} \, \frac{1}{k} \, \frac{\Gamma(E)\,\Gamma_0}{(E - 2\,\mu_B\,(B - B_\infty))^2 + \Gamma_{\mathrm{tot}}^2/4}.
\end{eqnarray}
Here, both inelastic loss and coupling to the continuum contribute to determining the total linewidth $\Gamma_{\mathrm{tot}} = \Gamma_0 + \Gamma(E)$ .
We assume that $\Gamma_0 = \hbar \, K_{\mathrm{ad}}\, (n_1 + n_2)$ where $K_{\mathrm{ad}}$ is the atom-dimer relaxation rate coefficient and $n_1$ ($n_2$) is the density of atoms in state $\left|1\right\rangle$ ($\left|2\right\rangle$).  For each state, the density decays according to $\dot{n} = -L_3 n^3$ since $L_2$ is itself proportional to density.  A simple expression for $L_3$ is obtained after performing a thermal average if we assume that the loss resonance is sharply peaked (i.e. $\Gamma_{\mathrm{tot}} \ll k_B T$) and the field is detuned from resonance such that $\Gamma(\mu_r\,(B - B_\infty)) \gg \Gamma_0$.  Under these conditions
\begin{eqnarray}
L_3 = \frac{2\,h^3 \, K_{\mathrm{ad}}}{(\pi m k_B T)^{3/2}} \, \exp\left[- \frac{\mu_r \, (B - B_\infty)}{k_B T} \right],
\label{eqn:L3}
\end{eqnarray}
which has a $1/e$ width given by $k_B\,T/\mu_r$.

We have fit both the $3.2\,\mu{\mathrm{K}}$ and $10\,\mu{\mathrm{K}}$ data sets using Eq.~\ref{eqn:L3} with $K_{\mathrm{ad}}$ as the only free parameter (solid lines in Fig.~\ref{fig:InelasticLoss}).  To evaluate $L_3$ in Eq.~\ref{eqn:L3}, we use $B_\infty = 543.286\,{\mathrm{G}}$ (determined from interaction energy shifts described below), the experimentally measured temperature (insets in Fig.~\ref{fig:InelasticLoss}), and the unknown atom-dimer relaxation rate constant $K_{\mathrm{ad}}$.  Eqn.~\ref{eqn:L3} provides an excellent description of the $3.2\,\mu{\mathrm{K}}$ data and a good description of the $10\,\mu{\mathrm{K}}$ data for atom-dimer relaxation rate coefficients $K_{\mathrm{ad}} = 3(1)\times10^{-10}\,{\mathrm{cm}}^3/{\mathrm{s}}$ and $K_{\mathrm{ad}} = 2(1)\times10^{-10}\,{\mathrm{cm}}^3/{\mathrm{s}}$ respectively.  Here, the uncertainty in $K_{\mathrm{ad}}$ is due to the uncertainty in trap frequencies.  From the average of both fits and including our systematic uncertainty in atom number ($\simeq 30\%$) we determine $K_{\mathrm{ad}} = 2.5 \pm 1.9 \times10^{-10}\,{\mathrm{cm}}^3/{\mathrm{s}}$. This narrow resonance can potentially be used as an energy-selective ``knife'' for forced evaporative cooling~\cite{Clark09}.  A slight reduction in temperature observed in both data sets when $\mu_r\,(B - B_\infty) \gtrsim 2\,E_F$ (see Fig.~\ref{fig:InelasticLoss} insets) may be a result of such evaporation.

\begin{figure}
\begin{center}
\includegraphics[width=3.25in]{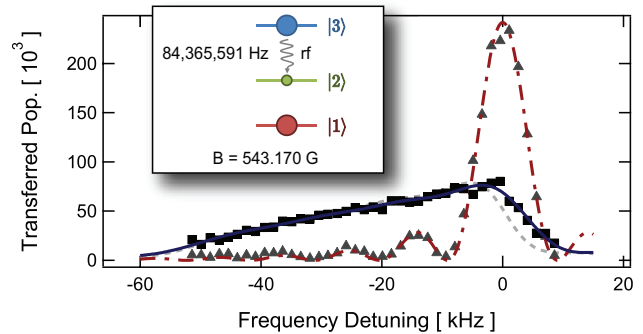}
\end{center}
\caption{\label{fig:RFLineshapes} Rabi spectra of the $\left|3 \right\rangle \rightarrow \left|2 \right\rangle$ transition for a $95\,\mu{\mathrm{s}}$--duration $\pi$--pulse.  ($\blacktriangle$) With no atoms in state $\left|1\right\rangle$, the spectrum accurately determines the $B$-field.  ($\blacksquare$)  With atoms in state $\left| 1 \right\rangle$ present, the lineshape is broadened and shifted by interactions.  The data is well-approximated (dashed) by convolving the Rabi transition lineshape (dot-dashed) with the lineshape obtained for a thermal gas in a harmonic trap assuming a frequency shift proportional to the local density.   Fits which allow the Rabi and offset frequencies to vary (solid line) are used to determine the mean shift $\bar{\nu}$.}
\end{figure}

To measure the dependence of interaction shifts on field and temperature, we have performed rf spectroscopy between states $\left| 3 \right\rangle$ and $\left|2\right\rangle$ (Fig.~\ref{fig:RFLineshapes}). The presence of an incoherent population in state $\left| 1 \right\rangle$ shifts the $\left| 3 \right\rangle \rightarrow \left|2\right\rangle$ transition frequency by the difference in interaction energy for a state-$\left|2\right\rangle$ or a state-$\left| 3 \right\rangle$ atom interacting with the background density of atoms in state $\left|1 \right\rangle$~\cite{Ketterle03A}.   The two-body contribution to mean-field interactions is determined by the real part of the complex scattering amplitude, $f$, yielding a shift of the rf transition given by
\begin{eqnarray}
\Delta \nu = \frac{2\,\hbar}{m} \, \left( -\mathrm{Re}[f_{13}] + \mathrm{Re}[f_{12}] \right) \, n_1
\label{eqn:MeanFieldShift}
\end{eqnarray}
where $n_1$ is the density of atoms in state $\left|1 \right\rangle$.  Interactions between atoms in states $\left|1 \right\rangle$ and $\left| 3 \right\rangle$ can be described by a contact potential parameterized by a scattering length $a_{13}$ for which $-\mathrm{Re}[f_{13}] = a_{13}$ at low energy (i.e. $k\,a_{13} \ll 1$).  However, the narrow $\left|1\right\rangle-\left|2 \right\rangle$ - resonance produces a scattering amplitude with a real part that is energy dependent.  In this case, $f_{12} = 1/(k\,\cot\delta_{12} - i\,k)$ where $k\,\cot\delta_{12}$ is given by Eq.~\ref{eqn:KCotDelta}.  For a thermal gas at temperature $T$ with density $n_1$, the expected frequency shift is given by Eq.~\ref{eqn:MeanFieldShift} where $\mathrm{Re}[f_{12}]$ is replaced by its thermal average to yield a $T$-dependent frequency shift $\Delta \nu(n_1,T,B)$.  Since the density distribution is inhomogeneous in the trap, the rf-transition lineshape is both shifted and broadened~\footnote{The inhomogeneous density distribution for a thermal gas in a harmonic trap produces an rf lineshape given by $P(\nu) = \frac{2}{\sqrt{\pi}} \, \frac{1}{\nu_{\mathrm{max}}} \, \sqrt{\ln\left(\frac{\nu_{\mathrm{max}}}{\nu} \right)}$ where $\nu_{\mathrm{max}}$ is the maximum shift (at the peak density) and $0 \leq \nu \leq \nu_{\mathrm{max}}$.  This is convolved with the lineshape for a Rabi $\pi$-pulse.} relative to that observed when atoms in state $\left| 1 \right\rangle$ are absent (Fig.~\ref{fig:RFLineshapes}).  In order to avoid inelastic loss and heating while adjusting $B$, we prepare a 50/50 mixture of atoms in states $\left|1 \right\rangle$ and $\left| 3 \right\rangle$ (rather than $\left|1 \right\rangle$ and $\left| 2 \right\rangle$), then shift to the field of interest and finally drive the rf transition from $\left|3 \right\rangle \rightarrow \left| 2 \right\rangle$ immediately before absorption imaging.

\begin{figure}
\begin{center}
\includegraphics[width=3.25in]{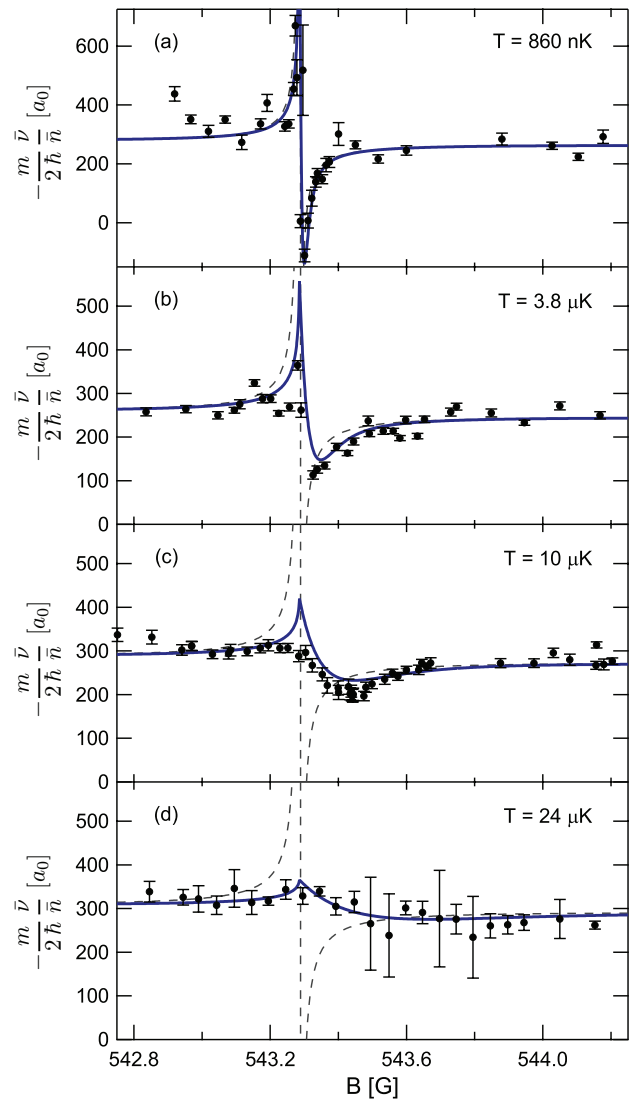}
\end{center}
\caption{\label{fig:IntMeas} Interaction-induced frequency shifts for clouds at different temperatures.  (dashed-lines) Shifts predicted if a contact potential describes $\left|1 \right\rangle - \left|2 \right\rangle$ collisions.  (solid lines) Shifts predicted by a mean-field theory that includes the energy dependence of the scattering phase shift expected near a narrow resonance.}
\end{figure}

In Fig.~\ref{fig:IntMeas} we plot $-\frac{m}{2\,\hbar}\,\frac{\bar{\nu}}{\bar{n}}$ for gases at different absolute temperatures where $\bar{\nu}$ is the mean value of the observed lineshape (relative to the bare transition frequency for an isolated atom) and $\bar{n}$ is the average density.  In terms of the mean-field prediction, the plotted quantity equals $a_* \equiv \langle -\mathrm{Re}[f_{12}(a_{bg},B,B_\infty,\Delta)] \rangle_{T} - a_{13}$.  Here $\langle ... \rangle_T$ indicates a thermal average.  The lowest temperature data (Fig.~\ref{fig:IntMeas}(a)) is fit to $\mathcal{A} \times a_*(T,a_{13},a_{bg},B,B_\infty,\Delta)$ where the amplitude $\mathcal{A}$ accommodates for our uncertainty in the absolute number and trap frequencies.  To minimize $\chi^2$, the amplitude $\mathcal{A}$, resonance location $B_\infty$, and width $\Delta$ are varied while the measured temperature $T$ and the slowly varying scattering lengths $a_{13}(B) \simeq -267\,a_0$ and $a_{bg}(B) \simeq 62\,a_0$~\cite{Julienne05,*JuliennePrivate} are fixed.  We find the minimum $\chi^2$ for $B_\infty = 543.286(3)\,{\mathrm{G}}$ and $\Delta = 0.10(1)\,{\mathrm{G}}$ where the quoted errors include the uncertainty in field calibration.

Using $B_\infty = 543.286\,{\mathrm{G}}$ and $\Delta = 0.10\,{\mathrm{G}}$ determined from the lowest temperature data, we plot $\mathcal{A} \times a_*$ in Figs.~\ref{fig:IntMeas}(a-d) for each specified temperature where only $\mathcal{A}$ is allowed to vary (solid lines).  For comparison, the prediction for a contact potential with $-{\mathrm{Re}}[f_{12}] = a_{bg} (1 - \Delta/(B - B_\infty))$ is also shown (dashed lines).  Clearly, using a contact interaction to model $\left|1 \right\rangle - \left|2 \right\rangle$-scattering fails to describe the observed interaction shifts as the temperature is increased.  The mean-field predictions which assume an energy-dependent $\mathrm{Re}[f_{12}]$ provide a much better description of the data.  In particular, the increase of the lineshape asymmetry with temperature, the behavior of the interaction energy on the high-field side of the resonance and the field at which the strongest attractive interaction energy is observed are well-described by this model.  However, even the energy-dependent model does not describe the data for fields extremely close to resonance (e.g. the $T = 10\,\mu{\mathrm{K}}$ data in Fig.~\ref{fig:IntMeas}(c)).  These discrepancies may be due to correlations in the many-body wavefunction not described by mean-field theory.  A recent calculation of the interaction energy near a narrow FR using a virial expansion (applicable for large $T/T_F$) also predicts asymmetric behavior across the resonance but does not quantitatively describe our data~\cite{Ho11}.

In summary, we demonstrated that the inelastic loss of fermions near a narrow FR is enhanced when the molecular state is above threshold, contrary to the behavior observed near broad resonances. We also demonstrated that interatomic interactions near a narrow FR cannot be described by a contact potential.  A mean-field theory which includes the full energy dependence of the scattering phase shift expected for a narrow FR provides a good, though not perfect, description of the observed interaction shifts.   This FR can be used to study strongly-correlated Fermi systems with energy-dependent interactions, potentially allowing for superfluids with the highest critical temperature ever achieved~\cite{Kokkelmans04,Ho11}, the observation of a breached-pair superfluid phase~\cite{Wilczek05}, and the determination of the equation of state for neutron star matter at densities comparable to or higher than the neutron drip density~\cite{Pethick05}.

\begin{acknowledgments}
We gratefully acknowledge enlightening discussions with C. Chin, X. Ciu, K. Gibble, V. Gurarie, T.-L. Ho, P. Julienne, S. Kokkelmans, and E. Timmermans regarding this work. This research was supported by the AFOSR (grant no. FA9550-08-1-0069), the NSF (grant no. PHY10-11156), and the ARO (grant no. W911NF-06-1-0398 which included  partial funding from the DARPA OLE program).
\end{acknowledgments}


\begin{thebibliography}{42}%
\makeatletter
\providecommand \@ifxundefined [1]{%
 \@ifx{#1\undefined}
}%
\providecommand \@ifnum [1]{%
 \ifnum #1\expandafter \@firstoftwo
 \else \expandafter \@secondoftwo
 \fi
}%
\providecommand \@ifx [1]{%
 \ifx #1\expandafter \@firstoftwo
 \else \expandafter \@secondoftwo
 \fi
}%
\providecommand \natexlab [1]{#1}%
\providecommand \enquote  [1]{``#1''}%
\providecommand \bibnamefont  [1]{#1}%
\providecommand \bibfnamefont [1]{#1}%
\providecommand \citenamefont [1]{#1}%
\providecommand \href@noop [0]{\@secondoftwo}%
\providecommand \href [0]{\begingroup \@sanitize@url \@href}%
\providecommand \@href[1]{\@@startlink{#1}\@@href}%
\providecommand \@@href[1]{\endgroup#1\@@endlink}%
\providecommand \@sanitize@url [0]{\catcode `\\12\catcode `\$12\catcode
  `\&12\catcode `\#12\catcode `\^12\catcode `\_12\catcode `\%12\relax}%
\providecommand \@@startlink[1]{}%
\providecommand \@@endlink[0]{}%
\providecommand \url  [0]{\begingroup\@sanitize@url \@url }%
\providecommand \@url [1]{\endgroup\@href {#1}{\urlprefix }}%
\providecommand \urlprefix  [0]{URL }%
\providecommand \Eprint [0]{\href }%
\providecommand \doibase [0]{http://dx.doi.org/}%
\providecommand \selectlanguage [0]{\@gobble}%
\providecommand \bibinfo  [0]{\@secondoftwo}%
\providecommand \bibfield  [0]{\@secondoftwo}%
\providecommand \translation [1]{[#1]}%
\providecommand \BibitemOpen [0]{}%
\providecommand \bibitemStop [0]{}%
\providecommand \bibitemNoStop [0]{.\EOS\space}%
\providecommand \EOS [0]{\spacefactor3000\relax}%
\providecommand \BibitemShut  [1]{\csname bibitem#1\endcsname}%
\let\auto@bib@innerbib\@empty
\bibitem [{\citenamefont {O'Hara}\ \emph
  {et~al.}(2002{\natexlab{a}})\citenamefont {O'Hara} \emph
  {et~al.}}]{Thomas02B}%
  \BibitemOpen
  \bibfield  {author} {\bibinfo {author} {\bibfnamefont {K.~M.}\ \bibnamefont
  {O'Hara}} \emph {et~al.},\ }\href@noop {} {\bibfield  {journal} {\bibinfo
  {journal} {Science}\ }\textbf {\bibinfo {volume} {298}},\ \bibinfo {pages}
  {2179} (\bibinfo {year} {2002}{\natexlab{a}})}\BibitemShut {NoStop}%
\bibitem [{\citenamefont {Bourdel}\ \emph {et~al.}(2003)\citenamefont {Bourdel}
  \emph {et~al.}}]{Salomon03}%
  \BibitemOpen
  \bibfield  {author} {\bibinfo {author} {\bibfnamefont {T.}~\bibnamefont
  {Bourdel}} \emph {et~al.},\ }\href@noop {} {\bibfield  {journal} {\bibinfo
  {journal} {Phys. Rev. Lett.}\ }\textbf {\bibinfo {volume} {91}},\ \bibinfo
  {pages} {020402} (\bibinfo {year} {2003})}\BibitemShut {NoStop}%
\bibitem [{\citenamefont {Regal}\ \emph {et~al.}(2004)\citenamefont {Regal},
  \citenamefont {Greiner},\ and\ \citenamefont {Jin}}]{Jin04}%
  \BibitemOpen
  \bibfield  {author} {\bibinfo {author} {\bibfnamefont {C.~A.}\ \bibnamefont
  {Regal}}, \bibinfo {author} {\bibfnamefont {M.}~\bibnamefont {Greiner}}, \
  and\ \bibinfo {author} {\bibfnamefont {D.~S.}\ \bibnamefont {Jin}},\
  }\href@noop {} {\bibfield  {journal} {\bibinfo  {journal} {Phys. Rev. Lett.}\
  }\textbf {\bibinfo {volume} {92}},\ \bibinfo {pages} {040403} (\bibinfo
  {year} {2004})}\BibitemShut {NoStop}%
\bibitem [{\citenamefont {Zwierlein}\ \emph {et~al.}(2004)\citenamefont
  {Zwierlein}, \citenamefont {Stan}, \citenamefont {Schunck}, \citenamefont
  {Raupach}, \citenamefont {Kerman},\ and\ \citenamefont
  {Ketterle}}]{Ketterle04}%
  \BibitemOpen
  \bibfield  {author} {\bibinfo {author} {\bibfnamefont {M.~W.}\ \bibnamefont
  {Zwierlein}}, \bibinfo {author} {\bibfnamefont {C.~A.}\ \bibnamefont {Stan}},
  \bibinfo {author} {\bibfnamefont {C.~H.}\ \bibnamefont {Schunck}}, \bibinfo
  {author} {\bibfnamefont {S.~M.~F.}\ \bibnamefont {Raupach}}, \bibinfo
  {author} {\bibfnamefont {A.~J.}\ \bibnamefont {Kerman}}, \ and\ \bibinfo
  {author} {\bibfnamefont {W.}~\bibnamefont {Ketterle}},\ }\href@noop {}
  {\bibfield  {journal} {\bibinfo  {journal} {Phys. Rev. Lett.}\ }\textbf
  {\bibinfo {volume} {92}},\ \bibinfo {pages} {120403} (\bibinfo {year}
  {2004})}\BibitemShut {NoStop}%
\bibitem [{\citenamefont {Bartenstein}\ \emph {et~al.}(2004)\citenamefont
  {Bartenstein}, \citenamefont {Altmeyer}, \citenamefont {Riedl}, \citenamefont
  {Jochim}, \citenamefont {Chin}, \citenamefont {Denschlag},\ and\
  \citenamefont {Grimm}}]{Grimm04}%
  \BibitemOpen
  \bibfield  {author} {\bibinfo {author} {\bibfnamefont {M.}~\bibnamefont
  {Bartenstein}}, \bibinfo {author} {\bibfnamefont {A.}~\bibnamefont
  {Altmeyer}}, \bibinfo {author} {\bibfnamefont {S.}~\bibnamefont {Riedl}},
  \bibinfo {author} {\bibfnamefont {S.}~\bibnamefont {Jochim}}, \bibinfo
  {author} {\bibfnamefont {C.}~\bibnamefont {Chin}}, \bibinfo {author}
  {\bibfnamefont {J.~H.}\ \bibnamefont {Denschlag}}, \ and\ \bibinfo {author}
  {\bibfnamefont {R.}~\bibnamefont {Grimm}},\ }\href@noop {} {\bibfield
  {journal} {\bibinfo  {journal} {Phys. Rev. Lett.}\ }\textbf {\bibinfo
  {volume} {92}},\ \bibinfo {pages} {203201} (\bibinfo {year}
  {2004})}\BibitemShut {NoStop}%
\bibitem [{\citenamefont {Zwierlein}\ \emph {et~al.}(2005)\citenamefont
  {Zwierlein}, \citenamefont {Schunck}, \citenamefont {Stan}, \citenamefont
  {Raupach},\ and\ \citenamefont {Ketterle}}]{Ketterle05}%
  \BibitemOpen
  \bibfield  {author} {\bibinfo {author} {\bibfnamefont {M.~W.}\ \bibnamefont
  {Zwierlein}}, \bibinfo {author} {\bibfnamefont {C.~H.}\ \bibnamefont
  {Schunck}}, \bibinfo {author} {\bibfnamefont {C.~A.}\ \bibnamefont {Stan}},
  \bibinfo {author} {\bibfnamefont {S.~M.~F.}\ \bibnamefont {Raupach}}, \ and\
  \bibinfo {author} {\bibfnamefont {W.}~\bibnamefont {Ketterle}},\ }\href@noop
  {} {\bibfield  {journal} {\bibinfo  {journal} {Phys. Rev. Lett.}\ }\textbf
  {\bibinfo {volume} {94}},\ \bibinfo {pages} {180401} (\bibinfo {year}
  {2005})}\BibitemShut {NoStop}%
\bibitem [{\citenamefont {Thomas}\ \emph {et~al.}(2005)\citenamefont {Thomas},
  \citenamefont {Kinast},\ and\ \citenamefont {Turlapov}}]{Thomas05A}%
  \BibitemOpen
  \bibfield  {author} {\bibinfo {author} {\bibfnamefont {J.~E.}\ \bibnamefont
  {Thomas}}, \bibinfo {author} {\bibfnamefont {J.}~\bibnamefont {Kinast}}, \
  and\ \bibinfo {author} {\bibfnamefont {A.}~\bibnamefont {Turlapov}},\
  }\href@noop {} {\bibfield  {journal} {\bibinfo  {journal} {Phys. Rev. Lett.}\
  }\textbf {\bibinfo {volume} {95}},\ \bibinfo {pages} {120402} (\bibinfo
  {year} {2005})}\BibitemShut {NoStop}%
\bibitem [{\citenamefont {Kinast}\ \emph {et~al.}(2005)\citenamefont {Kinast},
  \citenamefont {Turlapov}, \citenamefont {Thomas}, \citenamefont {Chen},
  \citenamefont {Stajic},\ and\ \citenamefont {Levin}}]{Thomas05B}%
  \BibitemOpen
  \bibfield  {author} {\bibinfo {author} {\bibfnamefont {J.}~\bibnamefont
  {Kinast}}, \bibinfo {author} {\bibfnamefont {A.}~\bibnamefont {Turlapov}},
  \bibinfo {author} {\bibfnamefont {J.}~\bibnamefont {Thomas}}, \bibinfo
  {author} {\bibfnamefont {Q.}~\bibnamefont {Chen}}, \bibinfo {author}
  {\bibfnamefont {J.}~\bibnamefont {Stajic}}, \ and\ \bibinfo {author}
  {\bibfnamefont {K.}~\bibnamefont {Levin}},\ }\href@noop {} {\bibfield
  {journal} {\bibinfo  {journal} {Science}\ }\textbf {\bibinfo {volume}
  {307}},\ \bibinfo {pages} {1296} (\bibinfo {year} {2005})}\BibitemShut
  {NoStop}%
\bibitem [{\citenamefont {Partridge}\ \emph {et~al.}(2006)\citenamefont
  {Partridge}, \citenamefont {Li}, \citenamefont {Kamar}, \citenamefont
  {Liao},\ and\ \citenamefont {Hulet}}]{Hulet06}%
  \BibitemOpen
  \bibfield  {author} {\bibinfo {author} {\bibfnamefont {G.}~\bibnamefont
  {Partridge}}, \bibinfo {author} {\bibfnamefont {W.}~\bibnamefont {Li}},
  \bibinfo {author} {\bibfnamefont {R.}~\bibnamefont {Kamar}}, \bibinfo
  {author} {\bibfnamefont {Y.}~\bibnamefont {Liao}}, \ and\ \bibinfo {author}
  {\bibfnamefont {R.}~\bibnamefont {Hulet}},\ }\href@noop {} {\bibfield
  {journal} {\bibinfo  {journal} {Science}\ }\textbf {\bibinfo {volume}
  {311}},\ \bibinfo {pages} {503} (\bibinfo {year} {2006})}\BibitemShut
  {NoStop}%
\bibitem [{Jin()}]{Jin06}%
  \BibitemOpen
  \href@noop {} {\bibfield  {journal} {\bibinfo  {journal} {Phys. Rev. Lett.}\
  }\textbf {\bibinfo {volume} {97}}}\BibitemShut {NoStop}%
\bibitem [{\citenamefont {Luo}\ \emph {et~al.}(2007)\citenamefont {Luo},
  \citenamefont {Clancy}, \citenamefont {Joseph}, \citenamefont {Kinast},\ and\
  \citenamefont {Thomas}}]{Thomas07}%
  \BibitemOpen
  \bibfield  {author} {\bibinfo {author} {\bibfnamefont {L.}~\bibnamefont
  {Luo}}, \bibinfo {author} {\bibfnamefont {B.}~\bibnamefont {Clancy}},
  \bibinfo {author} {\bibfnamefont {J.}~\bibnamefont {Joseph}}, \bibinfo
  {author} {\bibfnamefont {J.}~\bibnamefont {Kinast}}, \ and\ \bibinfo {author}
  {\bibfnamefont {J.~E.}\ \bibnamefont {Thomas}},\ }\href@noop {} {\bibfield
  {journal} {\bibinfo  {journal} {Phys. Rev. Lett.}\ }\textbf {\bibinfo
  {volume} {98}},\ \bibinfo {pages} {080402} (\bibinfo {year}
  {2007})}\BibitemShut {NoStop}%
\bibitem [{\citenamefont {Riedl}\ \emph {et~al.}(2008)\citenamefont {Riedl},
  \citenamefont {S\'anchez~Guajardo}, \citenamefont {Kohstall}, \citenamefont
  {Altmeyer}, \citenamefont {Wright}, \citenamefont {Denschlag}, \citenamefont
  {Grimm}, \citenamefont {Bruun},\ and\ \citenamefont {Smith}}]{Grimm08B}%
  \BibitemOpen
  \bibfield  {author} {\bibinfo {author} {\bibfnamefont {S.}~\bibnamefont
  {Riedl}}, \bibinfo {author} {\bibfnamefont {E.~R.}\ \bibnamefont
  {S\'anchez~Guajardo}}, \bibinfo {author} {\bibfnamefont {C.}~\bibnamefont
  {Kohstall}}, \bibinfo {author} {\bibfnamefont {A.}~\bibnamefont {Altmeyer}},
  \bibinfo {author} {\bibfnamefont {M.~J.}\ \bibnamefont {Wright}}, \bibinfo
  {author} {\bibfnamefont {J.~H.}\ \bibnamefont {Denschlag}}, \bibinfo {author}
  {\bibfnamefont {R.}~\bibnamefont {Grimm}}, \bibinfo {author} {\bibfnamefont
  {G.~M.}\ \bibnamefont {Bruun}}, \ and\ \bibinfo {author} {\bibfnamefont
  {H.}~\bibnamefont {Smith}},\ }\href@noop {} {\bibfield  {journal} {\bibinfo
  {journal} {Phys. Rev. A}\ }\textbf {\bibinfo {volume} {78}},\ \bibinfo
  {pages} {053609} (\bibinfo {year} {2008})}\BibitemShut {NoStop}%
\bibitem [{\citenamefont {Luo}\ and\ \citenamefont {Thomas}(2009)}]{Thomas09}%
  \BibitemOpen
  \bibfield  {author} {\bibinfo {author} {\bibfnamefont {L.}~\bibnamefont
  {Luo}}\ and\ \bibinfo {author} {\bibfnamefont {J.~E.}\ \bibnamefont
  {Thomas}},\ }\href@noop {} {\bibfield  {journal} {\bibinfo  {journal} {J. Low
  Temp. Phys.}\ }\textbf {\bibinfo {volume} {154}},\ \bibinfo {pages} {1}
  (\bibinfo {year} {2009})}\BibitemShut {NoStop}%
\bibitem [{\citenamefont {Nascimbene}\ \emph {et~al.}(2010)\citenamefont
  {Nascimbene}, \citenamefont {Navon}, \citenamefont {Jiang}, \citenamefont
  {Chevy},\ and\ \citenamefont {Salomon}}]{Salomon10}%
  \BibitemOpen
  \bibfield  {author} {\bibinfo {author} {\bibfnamefont {S.}~\bibnamefont
  {Nascimbene}}, \bibinfo {author} {\bibfnamefont {N.}~\bibnamefont {Navon}},
  \bibinfo {author} {\bibfnamefont {K.~J.}\ \bibnamefont {Jiang}}, \bibinfo
  {author} {\bibfnamefont {F.}~\bibnamefont {Chevy}}, \ and\ \bibinfo {author}
  {\bibfnamefont {C.}~\bibnamefont {Salomon}},\ }\href@noop {} {\bibfield
  {journal} {\bibinfo  {journal} {Nature}\ }\textbf {\bibinfo {volume} {463}},\
  \bibinfo {pages} {1057} (\bibinfo {year} {2010})}\BibitemShut {NoStop}%
\bibitem [{\citenamefont {Horikoshi}\ \emph {et~al.}(2010)\citenamefont
  {Horikoshi}, \citenamefont {Nakajima}, \citenamefont {Ueda},\ and\
  \citenamefont {Mukaiyama}}]{Mukaiyama10}%
  \BibitemOpen
  \bibfield  {author} {\bibinfo {author} {\bibfnamefont {M.}~\bibnamefont
  {Horikoshi}}, \bibinfo {author} {\bibfnamefont {S.}~\bibnamefont {Nakajima}},
  \bibinfo {author} {\bibfnamefont {M.}~\bibnamefont {Ueda}}, \ and\ \bibinfo
  {author} {\bibfnamefont {T.}~\bibnamefont {Mukaiyama}},\ }\href@noop {}
  {\bibfield  {journal} {\bibinfo  {journal} {Science}\ }\textbf {\bibinfo
  {volume} {327}},\ \bibinfo {pages} {442} (\bibinfo {year}
  {2010})}\BibitemShut {NoStop}%
\bibitem [{\citenamefont {Pieri}\ \emph {et~al.}(2011)\citenamefont {Pieri},
  \citenamefont {Perali}, \citenamefont {Strinati}, \citenamefont {Riedl},
  \citenamefont {Wright}, \citenamefont {Altmeyer}, \citenamefont {Kohstall},
  \citenamefont {S\'anchez~Guajardo}, \citenamefont {Hecker~Denschlag},\ and\
  \citenamefont {Grimm}}]{Grimm11B}%
  \BibitemOpen
  \bibfield  {author} {\bibinfo {author} {\bibfnamefont {P.}~\bibnamefont
  {Pieri}}, \bibinfo {author} {\bibfnamefont {A.}~\bibnamefont {Perali}},
  \bibinfo {author} {\bibfnamefont {G.~C.}\ \bibnamefont {Strinati}}, \bibinfo
  {author} {\bibfnamefont {S.}~\bibnamefont {Riedl}}, \bibinfo {author}
  {\bibfnamefont {M.~J.}\ \bibnamefont {Wright}}, \bibinfo {author}
  {\bibfnamefont {A.}~\bibnamefont {Altmeyer}}, \bibinfo {author}
  {\bibfnamefont {C.}~\bibnamefont {Kohstall}}, \bibinfo {author}
  {\bibfnamefont {E.~R.}\ \bibnamefont {S\'anchez~Guajardo}}, \bibinfo {author}
  {\bibfnamefont {J.}~\bibnamefont {Hecker~Denschlag}}, \ and\ \bibinfo
  {author} {\bibfnamefont {R.}~\bibnamefont {Grimm}},\ }\href@noop {}
  {\bibfield  {journal} {\bibinfo  {journal} {Phys. Rev. A}\ }\textbf {\bibinfo
  {volume} {84}},\ \bibinfo {pages} {011608} (\bibinfo {year}
  {2011})}\BibitemShut {NoStop}%
\bibitem [{\citenamefont {Gurarie}\ and\ \citenamefont
  {Radzihovsky}(2007)}]{Gurarie07}%
  \BibitemOpen
  \bibfield  {author} {\bibinfo {author} {\bibfnamefont {V.}~\bibnamefont
  {Gurarie}}\ and\ \bibinfo {author} {\bibfnamefont {L.}~\bibnamefont
  {Radzihovsky}},\ }\href {\doibase DOI: 10.1016/j.aop.2006.10.009} {\bibfield
  {journal} {\bibinfo  {journal} {Annals of Physics}\ }\textbf {\bibinfo
  {volume} {322}},\ \bibinfo {pages} {2 } (\bibinfo {year} {2007})}\BibitemShut
  {NoStop}%
\bibitem [{\citenamefont {Ketterle}\ and\ \citenamefont
  {Zwierlein}(2008)}]{Zwierlein08}%
  \BibitemOpen
  \bibfield  {author} {\bibinfo {author} {\bibfnamefont {W.}~\bibnamefont
  {Ketterle}}\ and\ \bibinfo {author} {\bibfnamefont {M.~W.}\ \bibnamefont
  {Zwierlein}},\ }\href@noop {} {\bibfield  {journal} {\bibinfo  {journal}
  {Riv. Nuovo Cimento}\ }\textbf {\bibinfo {volume} {31}},\ \bibinfo {pages}
  {247} (\bibinfo {year} {2008})}\BibitemShut {NoStop}%
\bibitem [{\citenamefont {Chin}\ \emph {et~al.}(2010)\citenamefont {Chin} \emph
  {et~al.}}]{Tiesinga10}%
  \BibitemOpen
  \bibfield  {author} {\bibinfo {author} {\bibfnamefont {C.}~\bibnamefont
  {Chin}} \emph {et~al.},\ }\href {\doibase 10.1103/RevModPhys.82.1225}
  {\bibfield  {journal} {\bibinfo  {journal} {Rev. Mod. Phys.}\ }\textbf
  {\bibinfo {volume} {82}},\ \bibinfo {pages} {1225} (\bibinfo {year}
  {2010})}\BibitemShut {NoStop}%
\bibitem [{\citenamefont {Ho}(2004)}]{Ho04}%
  \BibitemOpen
  \bibfield  {author} {\bibinfo {author} {\bibfnamefont {T.-L.}\ \bibnamefont
  {Ho}},\ }\href {\doibase 10.1103/PhysRevLett.92.090402} {\bibfield  {journal}
  {\bibinfo  {journal} {Phys. Rev. Lett.}\ }\textbf {\bibinfo {volume} {92}},\
  \bibinfo {pages} {090402} (\bibinfo {year} {2004})}\BibitemShut {NoStop}%
\bibitem [{\citenamefont {Ho}\ and\ \citenamefont {Mueller}(2004)}]{Mueller04}%
  \BibitemOpen
  \bibfield  {author} {\bibinfo {author} {\bibfnamefont {T.-L.}\ \bibnamefont
  {Ho}}\ and\ \bibinfo {author} {\bibfnamefont {E.~J.}\ \bibnamefont
  {Mueller}},\ }\href {\doibase 10.1103/PhysRevLett.92.160404} {\bibfield
  {journal} {\bibinfo  {journal} {Phys. Rev. Lett.}\ }\textbf {\bibinfo
  {volume} {92}},\ \bibinfo {pages} {160404} (\bibinfo {year}
  {2004})}\BibitemShut {NoStop}%
\bibitem [{\citenamefont {Carlson}\ \emph {et~al.}(2003)\citenamefont
  {Carlson}, \citenamefont {Chang}, \citenamefont {Pandharipande},\ and\
  \citenamefont {Schmidt}}]{Schmidt03}%
  \BibitemOpen
  \bibfield  {author} {\bibinfo {author} {\bibfnamefont {J.}~\bibnamefont
  {Carlson}}, \bibinfo {author} {\bibfnamefont {S.-Y.}\ \bibnamefont {Chang}},
  \bibinfo {author} {\bibfnamefont {V.~R.}\ \bibnamefont {Pandharipande}}, \
  and\ \bibinfo {author} {\bibfnamefont {K.~E.}\ \bibnamefont {Schmidt}},\
  }\href {\doibase 10.1103/PhysRevLett.91.050401} {\bibfield  {journal}
  {\bibinfo  {journal} {Phys. Rev. Lett.}\ }\textbf {\bibinfo {volume} {91}},\
  \bibinfo {pages} {050401} (\bibinfo {year} {2003})}\BibitemShut {NoStop}%
\bibitem [{\citenamefont {Palo}\ \emph {et~al.}(2004)\citenamefont {Palo},
  \citenamefont {Chiofalo}, \citenamefont {Holland},\ and\ \citenamefont
  {Kokkelmans}}]{Kokkelmans04}%
  \BibitemOpen
  \bibfield  {author} {\bibinfo {author} {\bibfnamefont {S.~D.}\ \bibnamefont
  {Palo}}, \bibinfo {author} {\bibfnamefont {M.}~\bibnamefont {Chiofalo}},
  \bibinfo {author} {\bibfnamefont {M.}~\bibnamefont {Holland}}, \ and\
  \bibinfo {author} {\bibfnamefont {S.}~\bibnamefont {Kokkelmans}},\ }\href
  {\doibase 10.1016/j.physleta.2004.05.034} {\bibfield  {journal} {\bibinfo
  {journal} {Physics Letters A}\ }\textbf {\bibinfo {volume} {327}},\ \bibinfo
  {pages} {490 } (\bibinfo {year} {2004})}\BibitemShut {NoStop}%
\bibitem [{\citenamefont {Bruun}\ and\ \citenamefont
  {Pethick}(2004)}]{Pethick04}%
  \BibitemOpen
  \bibfield  {author} {\bibinfo {author} {\bibfnamefont {G.~M.}\ \bibnamefont
  {Bruun}}\ and\ \bibinfo {author} {\bibfnamefont {C.~J.}\ \bibnamefont
  {Pethick}},\ }\href {\doibase 10.1103/PhysRevLett.92.140404} {\bibfield
  {journal} {\bibinfo  {journal} {Phys. Rev. Lett.}\ }\textbf {\bibinfo
  {volume} {92}},\ \bibinfo {pages} {140404} (\bibinfo {year}
  {2004})}\BibitemShut {NoStop}%
\bibitem [{\citenamefont {Schwenk}\ and\ \citenamefont
  {Pethick}(2005)}]{Pethick05}%
  \BibitemOpen
  \bibfield  {author} {\bibinfo {author} {\bibfnamefont {A.}~\bibnamefont
  {Schwenk}}\ and\ \bibinfo {author} {\bibfnamefont {C.~J.}\ \bibnamefont
  {Pethick}},\ }\href {\doibase 10.1103/PhysRevLett.95.160401} {\bibfield
  {journal} {\bibinfo  {journal} {Phys. Rev. Lett.}\ }\textbf {\bibinfo
  {volume} {95}},\ \bibinfo {pages} {160401} (\bibinfo {year}
  {2005})}\BibitemShut {NoStop}%
\bibitem [{\citenamefont {Ho}\ and\ \citenamefont {Cui}(2011)}]{Ho11}%
  \BibitemOpen
  \bibfield  {author} {\bibinfo {author} {\bibfnamefont {T.-L.}\ \bibnamefont
  {Ho}}\ and\ \bibinfo {author} {\bibfnamefont {X.}~\bibnamefont {Cui}},\
  }\href@noop {} {} (\bibinfo {year} {2011}),\ \Eprint
  {http://arxiv.org/abs/1105.4627} {arXiv:1105.4627} \BibitemShut {NoStop}%
\bibitem [{\citenamefont {Forbes}\ \emph {et~al.}(2005)\citenamefont {Forbes},
  \citenamefont {Gubankova}, \citenamefont {Liu},\ and\ \citenamefont
  {Wilczek}}]{Wilczek05}%
  \BibitemOpen
  \bibfield  {author} {\bibinfo {author} {\bibfnamefont {M.~M.}\ \bibnamefont
  {Forbes}}, \bibinfo {author} {\bibfnamefont {E.}~\bibnamefont {Gubankova}},
  \bibinfo {author} {\bibfnamefont {W.~V.}\ \bibnamefont {Liu}}, \ and\
  \bibinfo {author} {\bibfnamefont {F.}~\bibnamefont {Wilczek}},\ }\href
  {\doibase 10.1103/PhysRevLett.94.017001} {\bibfield  {journal} {\bibinfo
  {journal} {Phys. Rev. Lett.}\ }\textbf {\bibinfo {volume} {94}},\ \bibinfo
  {pages} {017001} (\bibinfo {year} {2005})}\BibitemShut {NoStop}%
\bibitem [{\citenamefont {Marcelis}\ and\ \citenamefont
  {Kokkelmans}(2006)}]{Kokkelmans06}%
  \BibitemOpen
  \bibfield  {author} {\bibinfo {author} {\bibfnamefont {B.}~\bibnamefont
  {Marcelis}}\ and\ \bibinfo {author} {\bibfnamefont {S.}~\bibnamefont
  {Kokkelmans}},\ }\href {\doibase 10.1103/PhysRevA.74.023606} {\bibfield
  {journal} {\bibinfo  {journal} {Phys. Rev. A}\ }\textbf {\bibinfo {volume}
  {74}},\ \bibinfo {pages} {023606} (\bibinfo {year} {2006})}\BibitemShut
  {NoStop}%
\bibitem [{\citenamefont {Marcelis}\ \emph {et~al.}(2008)\citenamefont
  {Marcelis}, \citenamefont {Verhaar},\ and\ \citenamefont
  {Kokkelmans}}]{Kokkelmans08}%
  \BibitemOpen
  \bibfield  {author} {\bibinfo {author} {\bibfnamefont {B.}~\bibnamefont
  {Marcelis}}, \bibinfo {author} {\bibfnamefont {B.}~\bibnamefont {Verhaar}}, \
  and\ \bibinfo {author} {\bibfnamefont {S.}~\bibnamefont {Kokkelmans}},\
  }\href {\doibase 10.1103/PhysRevLett.100.153201} {\bibfield  {journal}
  {\bibinfo  {journal} {Phys. Rev. Lett.}\ }\textbf {\bibinfo {volume} {100}},\
  \bibinfo {pages} {153201} (\bibinfo {year} {2008})}\BibitemShut {NoStop}%
\bibitem [{\citenamefont {Gupta}\ \emph {et~al.}(2003)\citenamefont {Gupta}
  \emph {et~al.}}]{Ketterle03A}%
  \BibitemOpen
  \bibfield  {author} {\bibinfo {author} {\bibfnamefont {S.}~\bibnamefont
  {Gupta}} \emph {et~al.},\ }\href {\doibase 10.1126/science.1085335}
  {\bibfield  {journal} {\bibinfo  {journal} {Science}\ }\textbf {\bibinfo
  {volume} {300}},\ \bibinfo {pages} {1723} (\bibinfo {year}
  {2003})}\BibitemShut {NoStop}%
\bibitem [{\citenamefont {Regal}\ and\ \citenamefont {Jin}(2003)}]{Jin03B}%
  \BibitemOpen
  \bibfield  {author} {\bibinfo {author} {\bibfnamefont {C.~A.}\ \bibnamefont
  {Regal}}\ and\ \bibinfo {author} {\bibfnamefont {D.~S.}\ \bibnamefont
  {Jin}},\ }\href {\doibase 10.1103/PhysRevLett.90.230404} {\bibfield
  {journal} {\bibinfo  {journal} {Phys. Rev. Lett.}\ }\textbf {\bibinfo
  {volume} {90}},\ \bibinfo {pages} {230404} (\bibinfo {year}
  {2003})}\BibitemShut {NoStop}%
\bibitem [{\citenamefont {O'Hara}\ \emph
  {et~al.}(2002{\natexlab{b}})\citenamefont {O'Hara} \emph
  {et~al.}}]{Thomas02A}%
  \BibitemOpen
  \bibfield  {author} {\bibinfo {author} {\bibfnamefont {K.~M.}\ \bibnamefont
  {O'Hara}} \emph {et~al.},\ }\href {\doibase 10.1103/PhysRevA.66.041401}
  {\bibfield  {journal} {\bibinfo  {journal} {Phys. Rev. A}\ }\textbf {\bibinfo
  {volume} {66}},\ \bibinfo {pages} {041401} (\bibinfo {year}
  {2002}{\natexlab{b}})}\BibitemShut {NoStop}%
\bibitem [{\citenamefont {Strecker}\ \emph {et~al.}(2003)\citenamefont
  {Strecker}, \citenamefont {Partridge},\ and\ \citenamefont
  {Hulet}}]{Hulet03}%
  \BibitemOpen
  \bibfield  {author} {\bibinfo {author} {\bibfnamefont {K.~E.}\ \bibnamefont
  {Strecker}}, \bibinfo {author} {\bibfnamefont {G.~B.}\ \bibnamefont
  {Partridge}}, \ and\ \bibinfo {author} {\bibfnamefont {R.~G.}\ \bibnamefont
  {Hulet}},\ }\href {\doibase 10.1103/PhysRevLett.91.080406} {\bibfield
  {journal} {\bibinfo  {journal} {Phys. Rev. Lett.}\ }\textbf {\bibinfo
  {volume} {91}},\ \bibinfo {pages} {080406} (\bibinfo {year}
  {2003})}\BibitemShut {NoStop}%
\bibitem [{\citenamefont {Wille}\ \emph {et~al.}(2008)\citenamefont {Wille}
  \emph {et~al.}}]{Grimm08A}%
  \BibitemOpen
  \bibfield  {author} {\bibinfo {author} {\bibfnamefont {E.}~\bibnamefont
  {Wille}} \emph {et~al.},\ }\href@noop {} {\bibfield  {journal} {\bibinfo
  {journal} {Phys. Rev. Lett.}\ }\textbf {\bibinfo {volume} {100}},\ \bibinfo
  {pages} {053201} (\bibinfo {year} {2008})}\BibitemShut {NoStop}%
\bibitem [{\citenamefont {Costa}\ \emph {et~al.}(2010)\citenamefont {Costa}
  \emph {et~al.}}]{Dieckmann10}%
  \BibitemOpen
  \bibfield  {author} {\bibinfo {author} {\bibfnamefont {L.}~\bibnamefont
  {Costa}} \emph {et~al.},\ }\href {\doibase 10.1103/PhysRevLett.105.123201}
  {\bibfield  {journal} {\bibinfo  {journal} {Phys. Rev. Lett.}\ }\textbf
  {\bibinfo {volume} {105}},\ \bibinfo {pages} {123201} (\bibinfo {year}
  {2010})}\BibitemShut {NoStop}%
\bibitem [{\citenamefont {Huckans}\ \emph {et~al.}(2009)\citenamefont {Huckans}
  \emph {et~al.}}]{OHara09}%
  \BibitemOpen
  \bibfield  {author} {\bibinfo {author} {\bibfnamefont {J.~H.}\ \bibnamefont
  {Huckans}} \emph {et~al.},\ }\href {\doibase 10.1103/PhysRevLett.102.165302}
  {\bibfield  {journal} {\bibinfo  {journal} {Phys. Rev. Lett.}\ }\textbf
  {\bibinfo {volume} {102}},\ \bibinfo {pages} {165302} (\bibinfo {year}
  {2009})}\BibitemShut {NoStop}%
\bibitem [{\citenamefont {Dieckmann}\ \emph {et~al.}(2002)\citenamefont
  {Dieckmann} \emph {et~al.}}]{Ketterle02}%
  \BibitemOpen
  \bibfield  {author} {\bibinfo {author} {\bibfnamefont {K.}~\bibnamefont
  {Dieckmann}} \emph {et~al.},\ }\href {\doibase 10.1103/PhysRevLett.89.203201}
  {\bibfield  {journal} {\bibinfo  {journal} {Phys. Rev. Lett.}\ }\textbf
  {\bibinfo {volume} {89}},\ \bibinfo {pages} {203201} (\bibinfo {year}
  {2002})}\BibitemShut {NoStop}%
\bibitem [{Note1()}]{Note1}%
  \BibitemOpen
  \bibinfo {note} {The geometric mean of the trap frequencies for the
  $3.2\protect \tmspace +\thinmuskip {.1667em}\mu {\protect \mathrm {K}}$
  ($10\protect \tmspace +\thinmuskip {.1667em}\mu {\protect \mathrm {K}}$) data
  is $\omega = 2\protect \tmspace +\thinmuskip {.1667em}\pi \times
  1.3(1)\protect \tmspace +\thinmuskip {.1667em}{\protect \mathrm {kHz}}$
  ($\omega = 2\protect \tmspace +\thinmuskip {.1667em}\pi \times 3.6(4)\protect
  \tmspace +\thinmuskip {.1667em}{\protect \mathrm {kHz}}$) and there are
  initially $N \simeq 3(1)\times 10^5$ ($N = 1.7(5) \times 10^5$) atoms in each
  spin state.}\BibitemShut {Stop}%
\bibitem [{\citenamefont {Mathey}\ \emph {et~al.}(2009)\citenamefont {Mathey}
  \emph {et~al.}}]{Clark09}%
  \BibitemOpen
  \bibfield  {author} {\bibinfo {author} {\bibfnamefont {L.}~\bibnamefont
  {Mathey}} \emph {et~al.},\ }\href {\doibase 10.1103/PhysRevA.80.030702}
  {\bibfield  {journal} {\bibinfo  {journal} {Phys. Rev. A}\ }\textbf {\bibinfo
  {volume} {80}},\ \bibinfo {pages} {030702} (\bibinfo {year}
  {2009})}\BibitemShut {NoStop}%
\bibitem [{Note2()}]{Note2}%
  \BibitemOpen
  \bibinfo {note} {The inhomogeneous density distribution for a thermal gas in
  a harmonic trap produces an rf lineshape given by $P(\nu ) = \protect \frac
  {2}{\protect \sqrt {\pi }} \protect \tmspace +\thinmuskip {.1667em} \protect
  \frac {1}{\nu _{\protect \mathrm {max}}} \protect \tmspace +\thinmuskip
  {.1667em} \protect \sqrt {\protect \qopname \relax o{ln}\left (\protect \frac
  {\nu _{\protect \mathrm {max}}}{\nu } \right )}$ where $\nu _{\protect
  \mathrm {max}}$ is the maximum shift (at the peak density) and $0 \leq \nu
  \leq \nu _{\protect \mathrm {max}}$. This is convolved with the lineshape for
  a Rabi $\pi $-pulse.}\BibitemShut {Stop}%
\bibitem [{\citenamefont {Bartenstein}\ \emph {et~al.}(2005)\citenamefont
  {Bartenstein} \emph {et~al.}}]{Julienne05}%
  \BibitemOpen
  \bibfield  {author} {\bibinfo {author} {\bibfnamefont {M.}~\bibnamefont
  {Bartenstein}} \emph {et~al.},\ }\href {\doibase
  10.1103/PhysRevLett.94.103201} {\bibfield  {journal} {\bibinfo  {journal}
  {Phys. Rev. Lett.}\ }\textbf {\bibinfo {volume} {94}},\ \bibinfo {pages}
  {103201} (\bibinfo {year} {2005})}\BibitemShut {NoStop}%
\bibitem [{Jul()}]{JuliennePrivate}%
  \BibitemOpen
  \href@noop {} {}\bibinfo {note} {P. Julienne (private
  communication).}\BibitemShut {Stop}%
\end{thebibliography}
\end{document}